\documentclass[conference]{IEEEtran}
\IEEEoverridecommandlockouts

\usepackage{amsmath,amssymb,amsfonts}
\usepackage{algorithmic}
\usepackage{graphicx}
\usepackage{textcomp}
\usepackage{xcolor}

\def\BibTeX{{\rm B\kern-.05em{\sc i\kern-.025em b}\kern-.08em
    T\kern-.1667em\lower.7ex\hbox{E}\kern-.125emX}}
\begin{document}

\title{DAIRHuM: A Platform for Directly Aligning AI Representations with Human Musical Judgments applied to Carnatic Music\\
}

\author{\IEEEauthorblockN{1\textsuperscript{st} Prashanth Thattai Ravikumar}
\IEEEauthorblockA{\textit{Department of Computing} \\
\textit{Goldsmiths, University of London}\\
London, UK \\
p.thattairavikumar@gold.ac.uk}
}

\maketitle

\begin{abstract}

Quantifying and aligning music AI model representations with human behavior is an important challenge in the field of MIR. This paper presents a platform for exploring the \textit{D}irect alignment between \textit{AI} music model \textit{R}epresentations and \textit{Hu}man \textit{M}usical judgments (\textit{DAIRHuM}). It is designed to enable musicians and experimentalists to label similarities in a dataset of music recordings, and examine a pre-trained model's alignment with their labels using quantitative scores and visual plots. \textit{DAIRHuM} is applied to analyze alignment between NSynth representations, and a rhythmic duet between two percussionists in a Carnatic quartet ensemble, an example of a genre where annotated data is scarce and assessing alignment is non-trivial. The results demonstrate significant findings on model alignment with human judgments of rhythmic harmony, while highlighting key differences in rhythm perception and music similarity judgments specific to Carnatic music. This work is among the first efforts to enable users to explore human-AI model alignment in Carnatic music and advance MIR research in Indian music while dealing with data scarcity and cultural specificity. The development of this platform provides greater accessibility to music AI tools for under-represented genres.

\end{abstract}


\section{Introduction}

Aligning AI models to human judgments is crucial for music information retrieval (MIR) tasks involving AI-generated music, as well as musical analysis. As AI systems are increasingly employed to generate, classify, and interpret music, ensuring their outputs resonate with human perception and preferences, is an important and non-trivial challenge. 


Research on aligning AI models with human perception in music has shown promising advancements in the fields of music expectation and surprise prediction. Masclef et al. demonstrated the utility of deep generative models, including diffusion models, in estimating surprise or "surprisal" values that align with human enjoyment. Their work observed the \textit{Wundt effect}, where moderate levels of surprise tend to increase listener enjoyment, indicating that generative models can replicate aspects of human anticipatory responses to music~\cite{masclef2023}. Similarly, Hansen et al. analyzed how high-entropy tones—those with greater uncertainty—attract listeners' attention for extended periods, impacting how musical phrases are perceived in terms of completeness. This work highlights the role of anticipatory processing in the brain, showing that certain predictive models can capture the segmentation processes involved in auditory sequence perception~\cite{Hansen_2021}. Research in other genre-specific contexts has illustrated the challenges faced when models trained on Western data are applied to indigenous music forms with diverse renditions across performers~\cite{moysis2023}. There have also been studies on Indian music investigating the computational analysis of ragas and  rhythms~\cite{srinivasamurthy2014rhythm}. However, exploring the direct alignment between AI music system representations and human music judgments remains an open challenge.

The paper proposes to tackle this specific problem of creating an accessible platform to assess an AI model's alignment to human judgments. The platform was designed to address the specific use-case of enabling \textit{musicians and experimentalists to rate similarities in a dataset of music-tracks, upload a model, and examine the model's alignment with their ratings}. The results of using the platform to analyze Carnatic percussion duets are presented, and discussed within emerging trends at the intersections of MIR, cognitive science, and cultural music studies.

\section{Carnatic percussion music}

Carnatic music, one of the two primary classical music traditions of India, alongside Hindustani music, has complex rhythmic structures that are central to both musical and allied South Asian dance forms like Bharatanatyam. Developing a platform to demonstrate AI alignment through this genre serves as an excellent example for illustrating accessibility in Music Information Retrieval (MIR) for culturally intricate genres.



In a typical Carnatic quartet performance, there are four improvising performers on the stage, the vocalist, the violinist, the lead percussionist (Mridangam) and one or more secondary percussionists (Kanjira/Ghatam/Morsing). The vocalist performs the main melody and the violinist plays the accompaniment melody. The lead percussionist improvises relative to the melody (Vocal and the Violin) and the secondary percussionist typically provides accompaniment to the lead percussionist.

Despite the rich improvisational interactions in Carnatic quartet performances, there has been limited computational work focused on modeling alignment between different improvisers in Carnatic music. Some efforts have been made to capture Indian percussion patterns, with datasets like the Mridangam Stroke Dataset ~\cite{guthrie2005mridangam} and the Tabla Dataset ~\cite{rao2011tabla} which provide isolated strokes and rhythmic patterns for these percussion instruments. Computational techniques, such as those by Guedes et al.~\cite{guedes2018modeling} have focused on generating patterns of Carnatic rhythms. There have also been studies on rhythm analysis for extracting musically meaningful rhythm related information from improvised solo recordings of Mridangam playing ~\cite{srinivasamurthy2016data}. However, these do not address the adaptive interaction between the lead and the secondary percussionists in an ensemble performance.



This paper focuses on studying model alignment using the concept of rhythmic harmony between the secondary percussionist (Kanjirist), lead (Mridangist) and the rest of the quartet ensemble performance. In particular, the \textit{DAIRHuM} platform will be used to evaluate the alignment of AI-generated rhythmic embeddings with human expert judgments of rhythmic harmony between the lead (Mridangist) and the secondary (Kanjirist) percussionists. The rest of the paper describes the tools offered by the system and the procedure for evaluation.

\section{The \textit{DAIRHuM} system}

The \textit{DAIRHuM} system is a Python package designed to support users in exploring and interpreting human-model alignment in musical tasks. Available on GitHub\footnote{https://github.com/prashanthtr/DAIRHuM}, its source code provides tools for creating recordings and their variations, analyzing embeddings to explore semantic similarities, and evaluating model performance against human judgments. The system enables users to systematically assess and interpret alignment using the following procedure.






\begin{itemize} 

\item \textbf{Labeling source and variations:} In this step, users compile a set of audio tracks, designating some as source tracks and others as variations. The source-variation labels are assigned based on some musically meaningful attributes of each track perceived by the user, e.g., the degree of melodic harmony or rhythmic congruence or similarity.



\item \textbf{Analysing embeddings:} In the second stage, pre-trained models are used to generate embeddings for each music track. These embeddings are analyzed to test for statistically significant differences, such as differences in aspects like rhythm or harmony, between variations of the same musical excerpt as interpreted by the AI model.


\item \textbf{Performance evaluation against human judgments:} Finally, the AI model's predictions are compared against human expert ratings from the dataset to determine the degree of alignment, measuring how closely the model’s outputs reflected the expert musical judgments.



    

\end{itemize}

\subsection{Labeling source and variations}



Typically, when users utilize this system, they will first gather a collection of audio tracks and assign them musically meaningful labels. Some of these audio tracks are identified as sources, and the other tracks are identified as variations of the source track. Users will assign source-variation labels based on a semantically significant attribute of the music they perceived within the tracks, e.g., degree of rhythmic or melodic harmony or any music similarity. The audio files are organized as a source-variation numbering scheme and placed in "Recordings" folder under "Data". 

In this specific example, the dataset contains Carnatic percussion accompaniment, featuring Mridangam and Kanjira, for three distinct melodic excerpts. Each melodic excerpt includes an original recording and five synthetically generated variations. The synthetic variations differ in the degree of rhythmic harmony in the accompaniment played on the Kanjira. This data is sourced from an empirical study on gathering musical harmony ratings from Carnatic experts for synthetically generated Kanjira tracks using a hand-crafted generative model~\cite{ravikumar2014playing}.  

The tracks are systematically labeled, with the original track for song 1 designated as "R1-V0" and the variations sequentially named from "R1-V1" to "R1-V5". This labeling scheme was consistently applied across all three songs (R1, R2, and R3) to ensure uniformity in the analysis. 

\begin{table}[htbp]
\caption{User ratings for original and 5 variations}
\begin{center}
\begin{tabular}{|c|c|c|c|c|c|c|}
\hline
\textbf{Recording}& \textbf{V0} & \textbf{V1} & \textbf{V2} & \textbf{V3} & \textbf{V4} & \textbf{V5} \\
\hline
{R1}& M & M & A & M & M & M \\
\hline
{R2}& S & S & M & S & S & HT \\
\hline
{R3}& HT & S & S & M & S & S \\
\hline

\end{tabular}
\label{tab1}
\end{center}
\end{table}


Table \ref{tab1} shows the different labels assigned to the three source recordings and their variations by an expert musician. Based on expert judgment, three primary labels describe the degree of rhythmic congruence: "mostly" (M), "half the time" (HT), and "sometimes" (S). Additional labels can be included such as "always congruent" (A) and "never congruent" (N). These labels provide a foundation for further similarity assessment and model evaluation.


\subsection{Analysing embeddings}


To proceed with assessing alignments between the AI model predictions and human judgments, users will present the labeled audio tracks to pre-trained models, and gather judgments of "sameness" between the tracks based on the similarity between  their musical embeddings. In this study, the NSynth model, a widely recognized neural network for audio signal processing was used to generate embeddings from the dataset created in the previous step.



Here’s an example of a procedure offered by the system to examine embeddings of audio tracks in the Kanjira dataset:

\begin{itemize}
    
    \item \textbf{Generate embeddings:} First, for each track in the dataset, the NSynth model is used to generate audio embeddings. E.g., suppose there are audio tracks labeled as R1-V0 (original recording for song 1) and R1-V1 (first variation of that song), the system generates embeddings for each using the NSynth model.
    
    \item \textbf{Experiment with similarity metrics:} Users will experiment with a selection of distance metrics to compare embeddings and find the one that aligns most closely with their chosen labeling scheme or musical characteristic of interest, such as rhythmic harmony.     
    
    

    \item \textbf{Choose statistical tests:} Users choose a permutation test to evaluate whether the observed differences in embeddings are statistically significant. E.g., suppose there are audio tracks labeled as R1-V0 (original recording for song 1) and R1-V1 (first variation of that song), the permutation test compares whether R1-V0 to R1-V1 were generated from the same distribution. 

    \item \textbf{Pairwise comparisons:} Pairwise statistical comparisons are conducted between all variations of the same song to detect any statistically significant differences. 
            
    \item \textbf{Assessing 'sameness':} The p-values from pairwise comparisons are used to interpret whether the model labels two variations of a musical track as 
    same or different.
        \begin{itemize}
            \item Tracks with p-values $<$ 0.05 are labeled "distinguishable(D)" indicating that the model interprets these tracks as being generated from different distributions.
            \item Tracks with p-values $>=$ 0.05 are labeled "indistinguishable(I)" indicating that the model interprets these tracks as being generated from the same underlying distribution.
        \end{itemize}
        
\end{itemize}

To make this procedure accessible, users have the option to utilize two preset similarity metrics - the Maximum Mean Discrepancy (MMD), and the Wasserstein distance. These presets offer an accessible way to explore commonly-used similarity metrics through libraries like SciPy and other Python packages. For users seeking more control, the platform also allows detailed customization of distance metrics for optimal alignment. Adjustable settings include kernel functions (e.g., Radial Basis Function (RBF) kernel) and specific values, such as gamma (\(\gamma\)) set to \( \frac{1}{\text{median distance between embeddings}} \), which enhances sensitivity to subtle rhythmic nuances.

Table \ref{tab2} shows the result of applying this procedure to the Kanjira dataset. In this specific test, embeddings were created using a pre-trained NSynth model. A permutation test with 1000 iterations was applied using the MMD metric with an RBF kernel, and \(\gamma\) set to \(1/\text{median distance between embeddings}\) to control the sensitivity to subtle rhythmic differences.

\begin{table}[htbp]
\caption{Analyzing embeddings of Recording 3}
\begin{center}
\begin{tabular}{|c|c|c|c|c|c|c|}
\hline
\textbf{Recording}& \textbf{V0} & \textbf{V1} & \textbf{V2} & \textbf{V3} & \textbf{V4} & \textbf{V5} \\
\hline
V0     & I  & D  & D  & D  & D  & D  \\ \hline
V1     & D  & I  & D  & D  & D  & D  \\ \hline
V2     & D  & D  & I  & I  & D  & D  \\ \hline
V3     & D  & D  & I  & I  & D  & D  \\ \hline
V4     & D  & D  & D  & D  & I  & I  \\ \hline
V5     & D  & D  & D  & D  & I  & I  \\ \hline


\end{tabular}
\label{tab2}
\end{center}
\end{table}

\subsection{Performance evaluation against human judgments}

The system is designed to measure the degree of alignment between a model's representations and human user representations. Specifically, it compares the notion of sameness/distinguishability derived from embeddings with the notion of sameness or distinguishability as used by humans in musical judgment. Thus, a high degree of alignment means that the model represents these concepts in the embedding space comparable to how humans would perceive them.

The simplest measure of alignment involves finding the number of matches between the AI-generated labels with the expert-assigned labels. The matches are used to quantify the degree of alignment through an alignment score or percentage. For example, Table \ref{tab5} shows matches (M) and non-matches (NM) between human and model judgments, indicating an alignment score of 66.66\%. For datasets with multiple sources and variations, the alignment scores for each source can be averaged to find the average alignment score for a given dataset.

\begin{figure}[htbp]
\centering
        \includegraphics[width=0.4\linewidth]{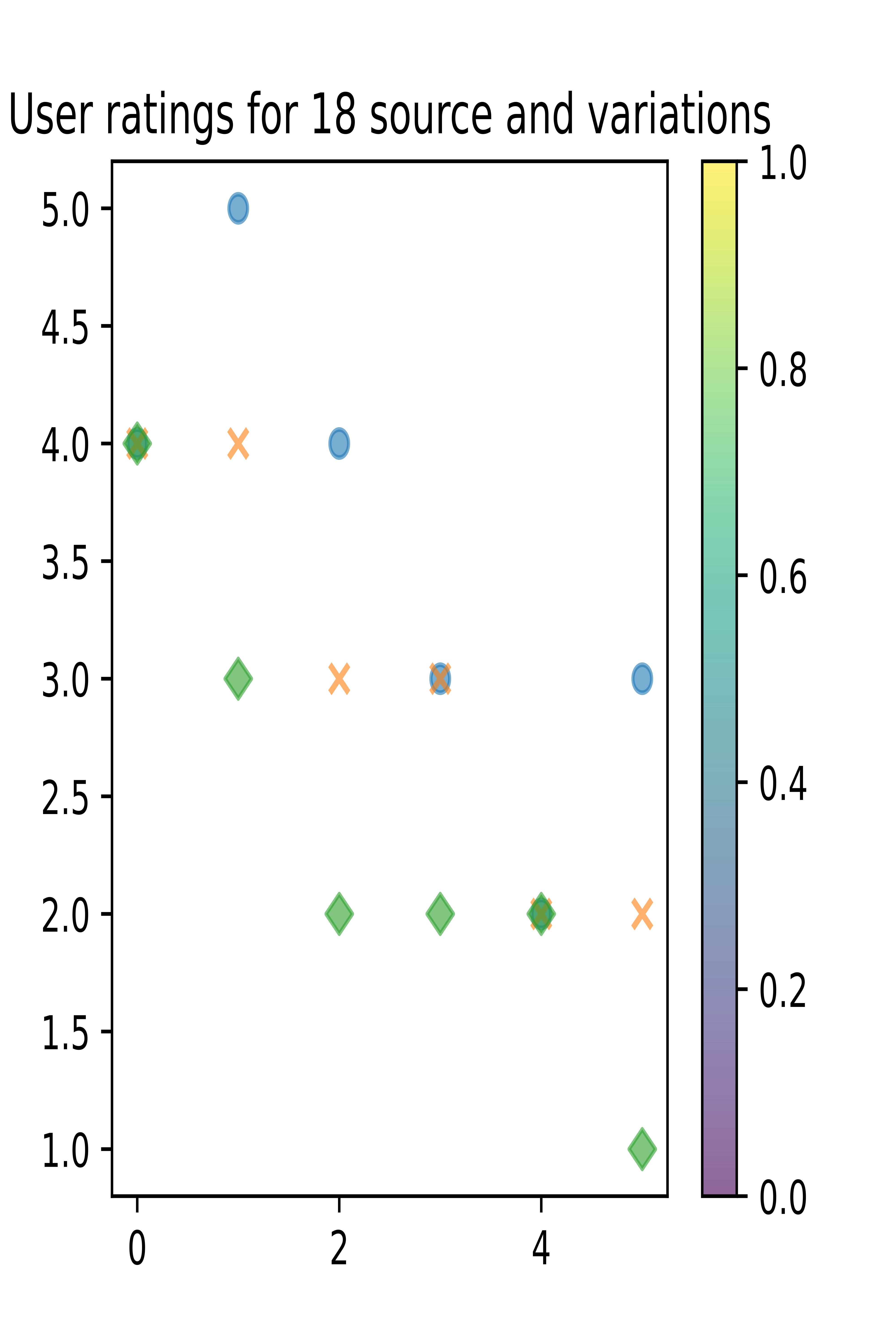} 
        \includegraphics[width=0.4\linewidth]{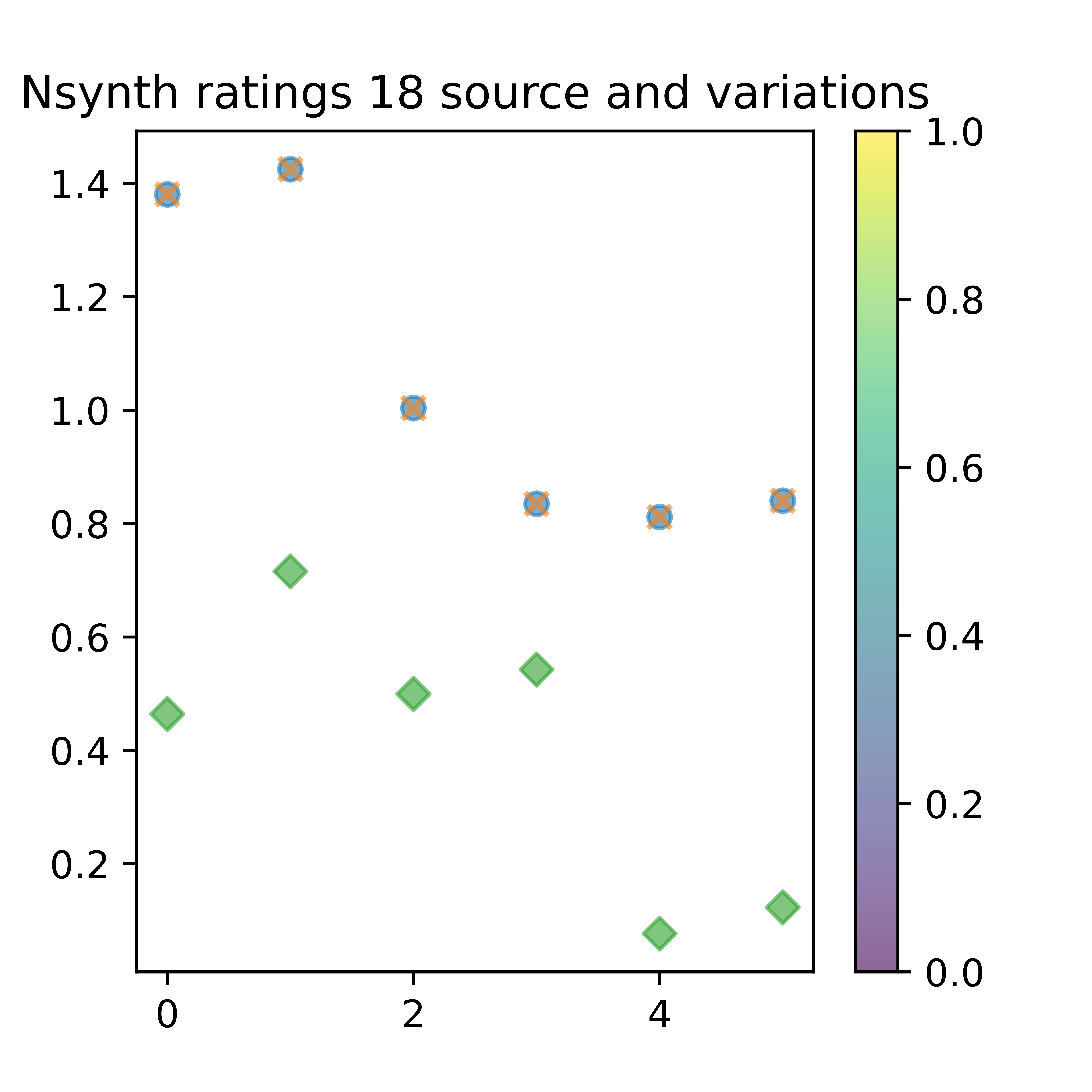} 
\caption{Comparing Human and NSynth's assessments of similarity}
\label{fig1}
\end{figure}

Additionally, the system generates plots offering a qualitative overview of the distribution of ratings within the representational space. Figure \ref{fig1} shows the overall trend indicating some similarities between the human and model judgments. Of particular interest are the regions of transition from high to low ratings, as these suggest plausible musical variations that induce categorical shifts in human perception or model interpretation. For instance, the plot of human ratings suggests that different musically varying accompaniments can have the same degree of harmony (3 Diamond shapes at Rating 2). On the other hand, for the same recordings, the model provides different ratings (Diamond shape at x-axis 4 is farther away). This suggests that there are qualitative differences between the expert's and the model's judgments, which could be explained through the subjectivity in the perception of certain rhythmic groupings.

\begin{table}[htbp]
\caption{Human vs Model ratings}
\begin{center}
\begin{tabular}{|c|c|c|c|c|c|c|}
\hline
\textbf{System/Human}& \textbf{V0} & \textbf{V1} & \textbf{V2} & \textbf{V3} & \textbf{V4} & \textbf{V5} \\
\hline
V0     & M   & M   & M   & M   & M   & M   \\ \hline
V1     & M   & M   & NM  & M   & NM  & NM  \\ \hline
V2     & M   & NM  & M   & NM  & NM  & NM  \\ \hline
V3     & M   & M   & NM  & M   & M   & M   \\ \hline
V4     & M   & NM  & NM  & M   & M   & M   \\ \hline
V5     & M   & NM  & NM  & M   & M   & M   \\ \hline
\multicolumn{7}{l}{$^{\mathrm{a}}$Comparing sameness assessments. M-Match, NM-No match}
\end{tabular}
\label{tab5}
\end{center}
\end{table}

\section{Discussion}


This paper presents a novel tool for Music Information Retrieval (MIR) researchers, musicians, and experimentalists to explore the direct alignment between AI-generated representations and human perceptual judgments. As a proof-of-concept, the platform was applied to analyze Carnatic percussion music, using embeddings generated by the NSynth model. Quantitative and qualitative results highlight similarities and differences in the human and the model's organization of rhythmic spaces, which presents opportunities to improve the context-sensitivity of the model. 

By providing a framework for evaluating how AI models interpret musical content in relation to expert human assessments, this platform offers a crucial advancement in the field of AI music alignment. While alignment has been studied in other domains, such as neural mechanisms and behavioral sciences, this work introduces a new avenue by directly addressing how AI representations of music align with human judgment in the context of musical genres, particularly those outside of the Western classical or popular domains. Platforms like Brain-score~\cite{schrimpf2020integrative}, which allow users to assess the alignment of neural models with human perception, have influenced this work. However, to the authors' knowledge, this is the first platform designed specifically to examine how AI models represent music in culturally nuanced genres, such as Carnatic percussion music, and compares these representations against expert human ratings. 


A key strength of the proposed system lies in its simplicity, as it only requires audio tracks, embeddings generated by a pre-trained model, and human labels. This bypasses the need for genre-specific representations and relies on a more universal approach through embeddings, which can be used to analyze any musical genre with human labels. This also makes it easier to apply the system to under-represented or low-resource musical genres, such as those from indigenous or non-Western cultures, such as Carnatic percussion music, which was used as an example in this paper. In this regard, \textit{DAIRHuM} offers a significant step forward in creating more inclusive, culture-driven MIR systems, and enables researchers and practitioners to assess AI models' representations across a diverse array of musical genres.


More specific to the context of Carnatic percussion music, the results reveal interesting differences in how AI models and human experts perceive rhythmic harmony. The clustering patterns observed between the human and model ratings suggest that while human experts group variations of similar rhythmic motifs together, the AI model tends to segregate these variations into more distinct, fine-grained categories. These findings prompt an important question concerning whether models, when trained on more culturally appropriate datasets, can evolve to organize musical variations in a manner that mirrors human experts’ perceptions. If so, it would represent a significant step toward creating AI systems for MIR, that not only align with human judgment but also develop a more nuanced understanding of cultural variations in music.

\section{Conclusion}

This paper presents a first-of-its-kind platform for directly assessing the alignment of AI models with human judgments. By analyzing a dataset of rhythmic duets in Carnatic percussion music with embedding techniques and pre-trained models, this provides a foundation for future research in culturally specific MIR. Furthermore, the insights gained from this platform underscore the complexities involved in modeling music similarity in a way that resonates with culturally informed human perception. This platform facilitates new possibilities for MIR applications in classification, generation, and music cognition studies within Indian as well other music genres.

\section*{Acknowledgment}

I would like to thank Lakshmi Narasimhan Govindarajan (ICoN Postdoctoral Fellow, MIT), for his insights on the development of this paper. 


\bibliographystyle{ieeetr}
\bibliography{refs}

\end{document}